\documentstyle[sprocl,graphicx]{article}

\newcommand{\half}{\frac{1}{2}}

\begin{document}
\title{HOW TO CALCULATE THE BUBBLE 
NUCLEATION RATE IN 
FIRST ORDER TRANSITIONS 
NON-PERTURBATIVELY$^{*}$\footnotetext{$^*$Presented by A. Tranberg at the Strong and Electroweak Matter 2000 in Marseilles.}}

\author{GUY D. MOORE}

\address{Department of physics, University of Washington, 
Seattle  WA 98195-1560 USA}

\author{KARI RUMMUKAINEN$^a$ and ANDERS TRANBERG$^b$$^{**}$\footnotetext{$^{**}$Currently at the Institute of Theoretical Physics, University of Amsterdam.}}

\address{$^a$NORDITA and $^b$Niels Bohr Institute, 
Blegdamsvej 17, DK-2100 Copenhagen \O, Denmark}



\maketitle
\abstracts{%
We present a new method for calculating the 
bubble nucleation rate in first order phase transitions
non-perturbatively on the lattice.  The method takes into 
account all fluctuations and
the full dynamical pre-factor. We also present results from applying
it to the cubic anisotropy model, which has a radiatively induced,
strongly first order phase transition.}


\noindent
Electroweak baryogenesis in the Standard Model (or extensions of it)
happens on or near the surface of the bubbles which nucleate and
subsequently grow during the first order Electroweak 
phase transition.\cite{NeKaCo}  
In this case, it is very difficult to compute the
bubble nucleation rate analytically with sufficient accuracy: since
the transition is radiatively generated, no classical bubble solution
exists, and the whole idea of separating the classical bubble from 
the fluctuation
determinant, needed for Langer's nucleation theory,\cite{Lan} becomes
cumbersome.  Furthermore, the long-distance physics of the Electroweak
theory is inherently non-perturbative, making lattice simulations necessary.


How can one calculate the nucleation rate on the lattice?  The most
straightforward method is to take an ensemble of configurations in the
meta\-stable state, evolve each with Hamilton's equations of motion,
and wait for tunneling to happen. This has been done, for instance
in the Ising model\,\cite{Isi} and the $\phi^{4}$ model.\cite{phi4rth}  
However, the cooling rate of the Universe during the
Electroweak phase transition is many orders of magnitude smaller
than the timescale of microscopic interactions. 
Thus, the system
has plenty of time for probing its phase space, and the tunneling will
happen through very strongly suppressed configurations ($p \sim e^{-100}$)
after a small amount of supercooling.  However, it is impossible to make
Monte Carlo simulations with $e^{100}$ iterations!


Below, we shall describe a Monte Carlo method which fully overcomes
this problem, and which can be used to calculate both the static and
the dynamical parts of the nucleation rate.
Instead of using the full $SU(2)\times
U(1)+$Higgs, we chose as our toy model the cubic anisotropy model,
which has the desired features:

$\bullet$ It has a radiatively induced, first order transition.  As mentioned
above, this makes analytical computations quite difficult, see ref.\cite{tetradis}

$\bullet$ It is simple to simulate on the lattice.

\noindent
The continuum Hamiltonian of the cubic anisotropy model in three
dimensions is ($\phi_{1}$ and $\phi_{2}$ are scalar 
fields, $\pi_1$ and $\pi_2$ the
associated canonical momenta):
\begin{equation}
{\mathcal{H}}=\sum_{a=1,2} \bigg[ 
\half \pi_a^2 + 
\half \partial_{i}\phi_{a}\partial_{i}\phi_{a} +
\half m^{2}\phi_a^{2} +
\frac{\lambda_{1}}{24}\phi_a^{4}
\bigg]
+\frac{\lambda_{2}}{4}\phi_{1}^{2}\phi_{2}^{2} \label{ham}
\end{equation}
The Hamiltonian is discretized on the lattice using an improved
(next-to-nearest neighbour) Laplacian. Included also are loop
corrections to second order, which enter in the lattice versions of
$\lambda_{1}$, $\lambda_{2}$, $\phi_{1}$, $\phi_{2}$ and $m^{2}$. We
fix $\lambda_{2}=8\lambda_{1}$, giving us a strongly first order phase
transition between the symmetric and broken phases, as we vary
$m^{2}$.  In this case $m^{2}$ plays the role of the temperature. We
find the critical $m^{2}$, $m^{2}_{c}$, the latent heat and the
interface tension.
The full results will be presented in a future paper.\cite{MoRuTr}

The calculation of the rate can be split up into two main parts, the
static probability of the critical bubbles (part $I$), and the 
dynamical flux of the phase space through the 
bubble configurations (part $II$).
As an order parameter, we choose the space average of 
$\Phi^{2}=\phi_{1}^{2}+\phi_{2}^{2}$.

\begin{figure}
\begin{center}
\includegraphics[width=6cm]{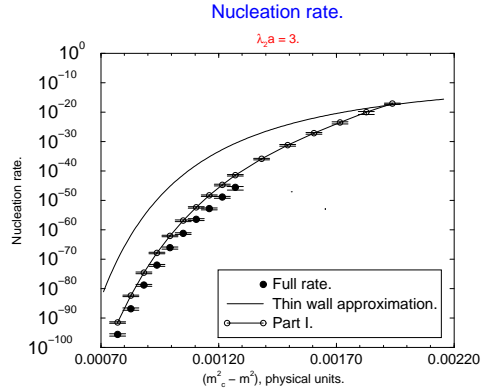}
\caption{The full nucleation rate compared 
to the thin-wall approximation (line).}
\label{rate}
\end{center}
\end{figure}

\begin{list}{$\bullet$}{\leftmargin=3mm}
\item 
The non-dynamical part $I$ is the probability of having a critical
bubble configuration. It is found by calculating the probability
distribution $P(\Phi^2)$ of the order parameter (see Fig.~\ref{probdist}),
and defining the critical bubble to be at the
minimum, $Z$. We define a region
$[Z-\frac{\epsilon}{2},Z+\frac{\epsilon}{2}]$ as the critical bubble
region.

At the critical temperature, because of the finiteness of the
lattice, the most suppressed configurations 
are slabs (see Fig.~\ref{probdist}),
and the bubble configurations are found on the inner
sides of the two peaks. Lowering the temperature ($m^{2}$), the
critical bubble becomes smaller and eventually it will fit inside the
lattice volume. Part I is then the probability density in the critical
bubble region ($\frac{P_{z}}{\epsilon}$) divided by the probability of
the metastable state ($P_{A}$).
\begin{eqnarray}
I &=&\frac{P_{z}}{P_{A}\epsilon}~~,
~~P_{Z}=\int_{Z-\frac{\epsilon}{2}}^{Z+\frac{\epsilon}{2}}P(\Phi^2)d\Phi^2~~,
~~P_{A}=\int_{0}^{Z}P(\Phi^2)d\Phi^2
\end{eqnarray}
$P(\Phi^2)$ is found by using multicanonical methods.\cite{MoRuTr}
It can be reweighted to several different 
values of $m^{2}$, and part $I(m^2)$ calculated.
\begin{figure}
\begin{center}
\begin{tabular}{c c}
\includegraphics[width=5.5cm]{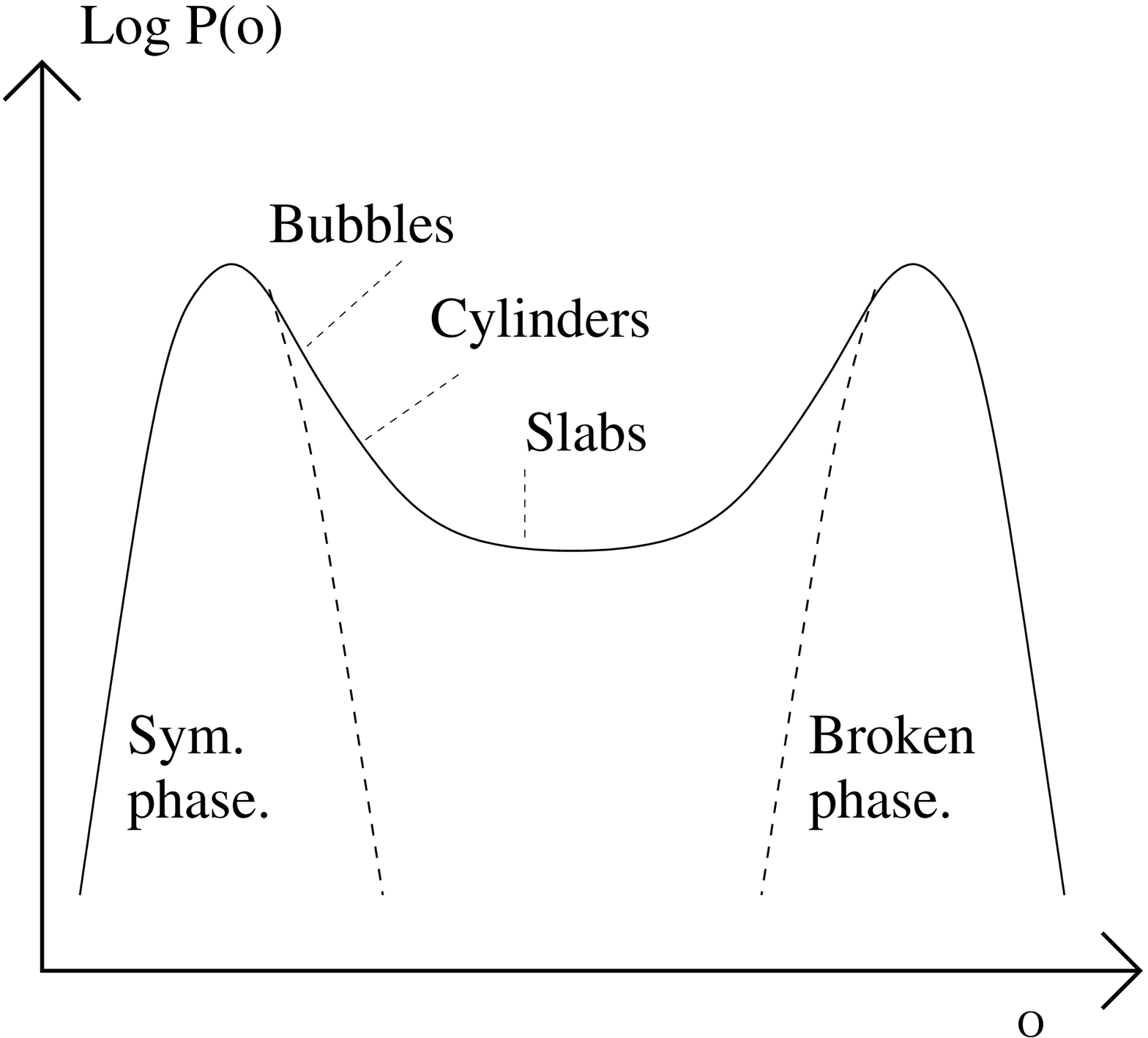}
&
\includegraphics[width=5.5cm]{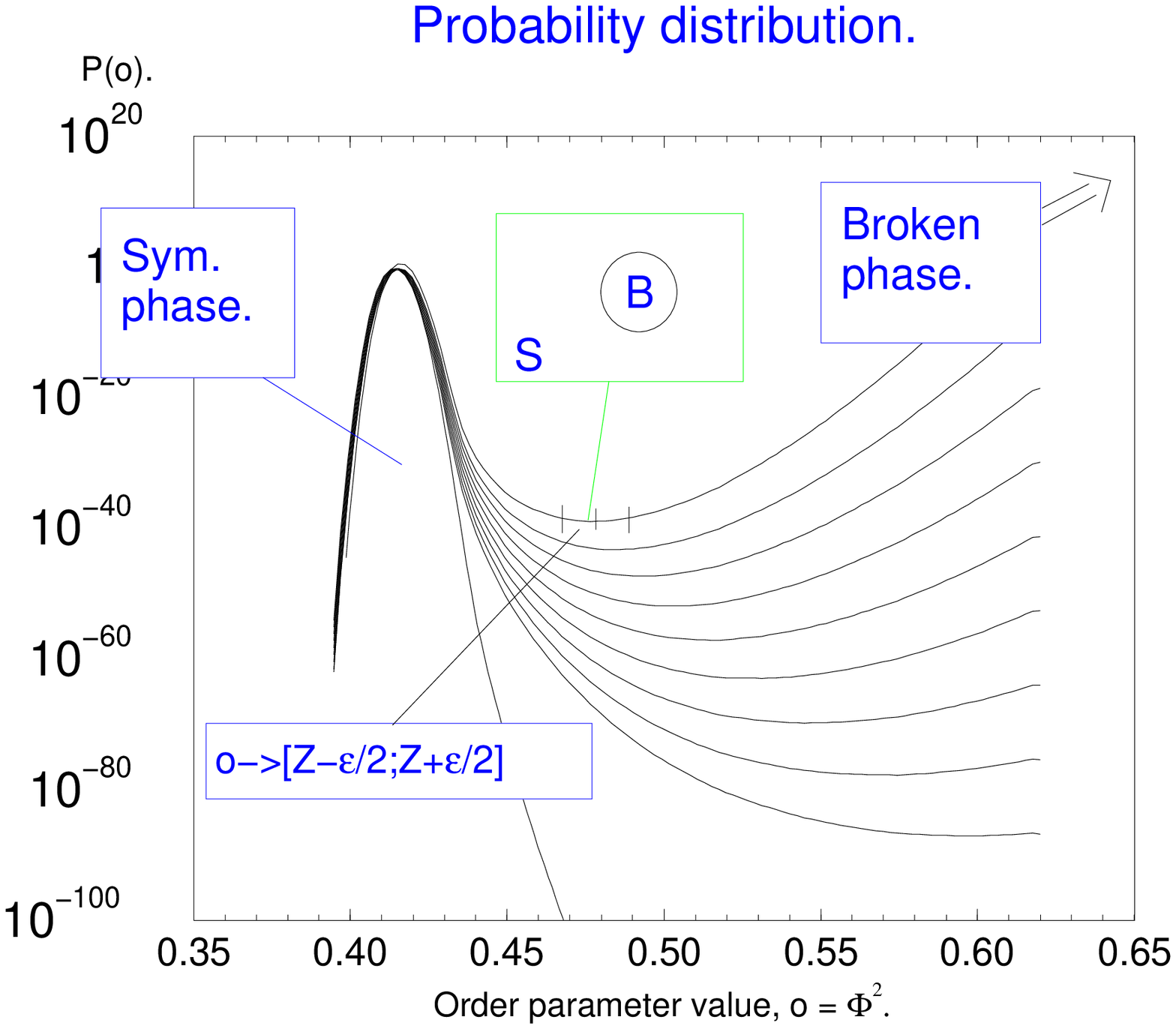}
\end{tabular}
\caption[a]{Sketch of the different geometries 
of the configurations in a finite
box (left).The probability distribution is reweighted to several
different values of $m^{2}$ where the critical bubble fits inside the
lattice volume. For each $m^{2}$, $Z$ is determined and $\epsilon$ is
chosen (right).}
\label{probdist}
\end{center}
\end{figure}
\item Part $IIa$ is the flux of order parameter 
through the critical bubble value, $Z$. 
It can be calculated analytically\cite{MoRuTr}:
\begin{equation}
IIa \equiv \frac{1}{2}\langle|\frac{d\Phi^{2}}{dt}|\rangle
~=~\sqrt{\frac{2Z}{\pi V}}
\end{equation}
\item
However, not all of the flux in $IIa$ leads to tunneling. The
bubble wobbles back and forth for some time before falling into one of
the minima.  Since we are actually sampling the {\em configurations\,}
at $\Phi^2=Z$,
we are over-counting the true phase space flux 
from the symmetric phase to the broken phase
by the average number of crossings of $Z$ the trajectories make.
Therefore, we use the Hamiltonian in Eq.~(\ref{ham}) to calculate the real time
trajectories forward and backwards in time from initial
configurations in the critical bubble region
$[Z-\frac{\epsilon}{2},Z+\frac{\epsilon}{2}]$. Gluing the two together
into a full trajectory, we can then count the number of crossings, and
determine whether the trajectory starts and ends in different
minima. 
%
%
Finally, we can correct the part $IIa$ with part $IIb$:
\begin{eqnarray}
IIb &\equiv &\frac{1}{N}\sum_{i}\frac{\delta_{i}}{c_{i}}
\end{eqnarray}
where $N$ is the number of trajectories, $c_{i}$ is the number of
crossings of the i'th trajectory and $\delta_{i}$ is 1 if the
trajectory tunneled, 0 if not. 

However, since the Hamiltonian evolution of the bubble conserves the
total energy, the finite size of the system poses a problem: when
the bubble grows (shrinks), it releases (absorbs) latent heat, and
the temperature grows (decreases).  This 
effect tends to unphysically ``stabilize''
the bubble around the critical radius, see Fig.~\ref{traj}.
This can be solved by going to very large volumes, which is too expensive,
or by adding a small amount of thermal noise in the system.
The thermal noise consists of an update of the momenta:
\begin{eqnarray}
\pi'=\sqrt{1-\epsilon^{2}}\pi+\epsilon \xi~~,~~\epsilon=\sqrt{\frac{2\delta
t}{L}}
\end{eqnarray}
where $\xi$ is taken from a Gaussian of width 1. The amplitude of the
noise is tuned so that the $\pi$'s are thermalized after evolution of
length $t=\lambda_{2}aL$. The final rate is insensitive to the precise
value of $\epsilon$.

\begin{figure}
\begin{center}
\includegraphics[width=8cm]{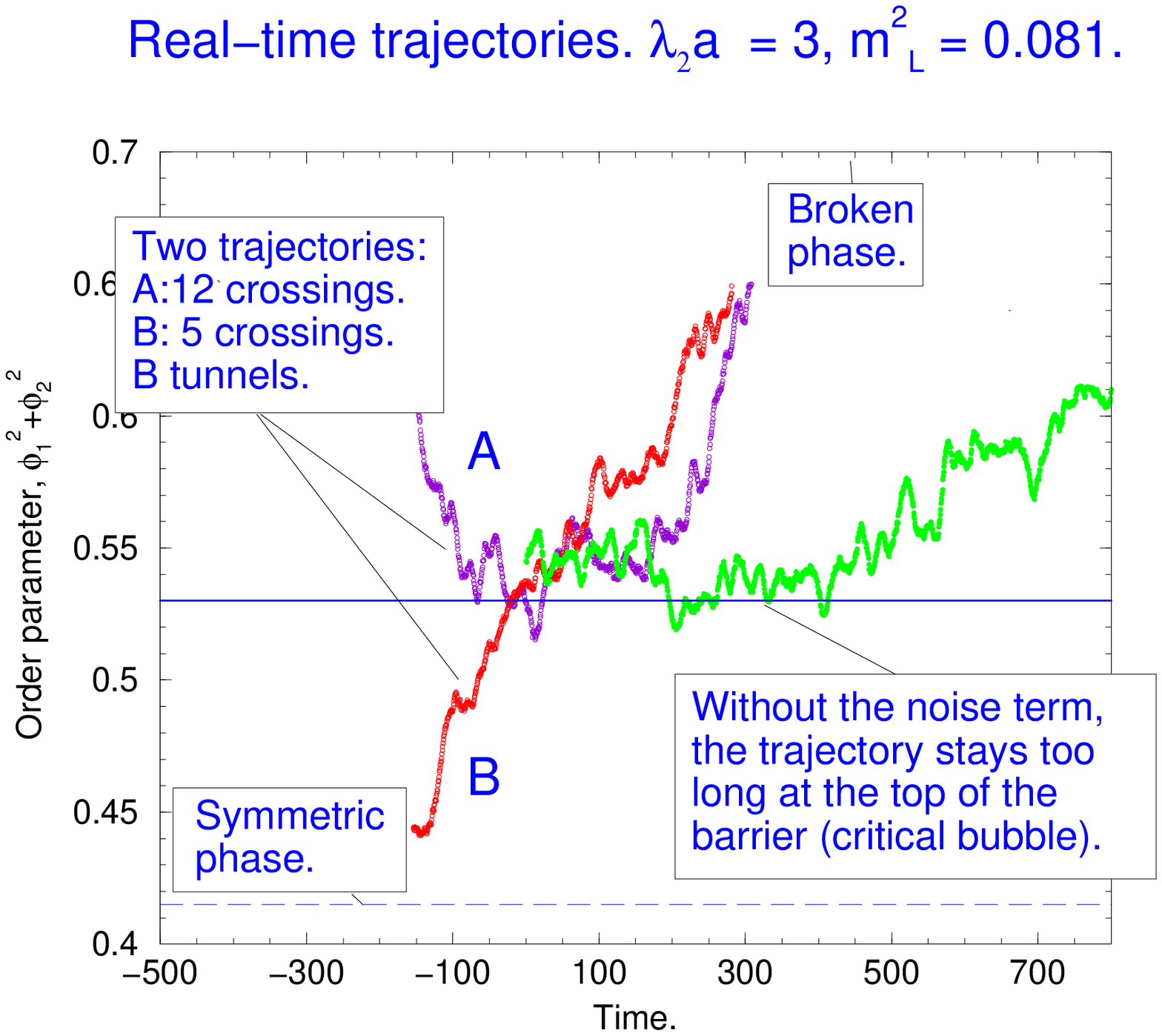}
\caption[a]{
From the trajectories we can count the crossings and see if it is a
tunneling trajectory. Without the noise term, the bubble stays
on the barrier too long and the number of crossings becomes too large.}
\label{traj}
\end{center}
\end{figure}

\end{list}
The final rate is (see Figure \ref{rate}):
\begin{equation}
\Gamma=\frac{I\times IIa\times IIb}{V}
\end{equation}
In Fig.~\ref{rate} we compare our result with the thin-wall approximation.
At smallest supercoolings we have studied, the thin-wall result gives a rate 
$\sim 25$ orders of magnitude too large. 
However, for a fixed rate, the thin-wall
approximation gives ``only'' $\sim 30$\% too small supercooling $\delta m^2$.

To conclude, we have described a lattice Monte Carlo method for
calculating the full bubble nucleation rate in first order phase
transitions.  The method is especially well suited for
transitions which happen through very strongly suppressed bubble
configurations; indeed, most first order phase transitions in
Nature fall into this class.  We have applied this method to the cubic
anisotropy model in 3D, for a full description we refer to the
forthcoming paper.\cite{MoRuTr}
This method has also been applied to $SU(2)$+Higgs theory, but with a
heat bath evolution instead of the Hamiltonian real time evolution.\cite{MoRu}

\end{document}